\RequirePackage{fix-cm}
\documentclass[twocolumn]{svjour3}          
\smartqed  
\usepackage{graphicx}
\mathchardef\bigtilde="0365

\def\la{\langle}
\def\ra{\rangle}
\def\s{\sigma}

\begin{document}


\title{Continuum limit of susceptibility from strong coupling expansion
}
\subtitle{Two dimensional non-linear $O(N)$ sigma model at $N\ge 3$}


\author{Hirofumi Yamada
}


\institute{H. Yamada \at
             Division of Mathematics and Science, Chiba Institute of Technology, 
\\Shibazono 2-1-1, Narashino, Chiba 275-0023, Japan\\
              \email{yamada.hirofumi@it-chiba.ac.jp}           
}

\date{Received: date / Accepted: date}

\maketitle

\begin{abstract}
Based on the strong coupling expansion, we reinvestigate the scaling behavior of the susceptibility $\chi$ of two-dimensional $O(N)$ sigma model on the square lattice by the use of Pad\'e-Borel approximants.  To exploit the Borel transform, we express the bare coupling $g$ in series expansion in $\chi$.    At large $N$, Pad\'e-Borel approximants exhibit the scaling behavior at the four-loop level.  Then, the estimation of the non-perturbative constant associated with the susceptibility is performed for $N\ge 3$ and the results are compared with the available theoretical results and Monte Carlo data.
\end{abstract}

\section{Introduction}
\label{intro}
In the lattice formulation of physical models, the lattice spacing $a$ is a basic parameter under feasible theoretical control.  It is therefore natural to describe observables in terms of $a$.   We concern here with the possibility of the use of large $a$ expansion for the approximation of the continuum scaling.

In a recent paper \cite{yam1}, continuum scaling of lattice field theories was studied from the view point of strong coupling expansion.  The model studied there is the non-linear $O(N)$ sigma model on two-dimensional square lattice.  This model enjoys properties similar to Yang-Mills model such as asymptotic freedom and dynamical mass generation  \cite{pol} and provides us a convenient testing ground of a new computational scheme.  Since the large lattice spacing means the large bare coupling due to asymptotic freedom, large $a$ expansion is equivalent with the strong couping expansion.   
It was then demonstrated that the Pad\'e-Borel approximants of the large $a$ ($a$: the lattice spacing) series of bare coupling $g$ shows four-loop scaling for large $N$.  The estimation of the non-perturbative constant $C_{\xi}$, which enters into the relation of the correlation length $\xi$ with $g$ such as $\xi=C_{\xi}\exp(\frac{2\pi N}{N-2})\big(\frac{2\pi N\beta}{N-2}\big)^{-\frac{1}{N-2}}(1+\cdots)$ where $\beta=1/g^2$, was then estimated at $N\ge 3$.  The result of estimation is in good agreement with the result via thermodynamic Bethe ansatz due to Hasenfratz et. al \cite{hasen}.

Let us briefly survey the outline of \cite{yam1}.  In \cite{yam1}, the basic variable by which the bare coupling is expressed was the momentum mass $M$ defined by the zero momentum limit of two-point correlation function.  $M$ is the practical replacement of $a$.   In the Fourier transformed form, $M$ appears in the correlation function as
\begin{equation}
\beta\la\vec{\sigma}_{\bf 0}\cdot \vec{ \sigma}_{\bf x}\ra=N\int_{-\pi}^{\pi}\frac{d^2 p}{(2\pi)^2}\frac{Z\exp(-i{\bf p}\cdot{\bf n})}{M+p^2+O(p^4)}.
\label{mass}
\end{equation}
$Z$ denotes the wave function renormalization and the coefficient of $p^2$ is rescaled to be one.  Conventionally, $M$ is considered as function of $\beta$.  In \cite{yam1}, however, another approach was employed to access the scaling behaviors:  Mutual roles of $\beta$ and $M$ was inverted and $\beta$ was expressed in $M$ in the form $\beta=\sum_{k=1}a_{k}M^{-k}$.  Then, recalling that one must consider the behavior of $\beta$ at $1/M\gg 1$,  Borel transform with respect to $M$ was attempted.   Borel transforming the relation giving $\bar\beta=\sum_{k=1}(a_{k}/k!)\bar M^{-k}$ (bar variables denote the Borel transformed ones) which is entire series and improving the series by Pad\'e approximants,  the non-perturbative constant $C_{\xi}$ has been estimated.   We would note that Borel transform in our use has connection with the scaling transformation, as illustrated in \cite{yam1}.

When the second order phase transition is under consideration, the choice of the momentum mass as a basic variable describing the system seems to be natural, since it spells the typical cooperative length of the system.   There is, however, another variable which may play a similar role with $1/M$ or $\xi$, the susceptibility $\chi$.   In this work, the susceptibility $\chi$ is defined as the sum of all two-point correlation functions times $\beta$,
\begin{equation}
\chi=\frac{\beta}{N}\sum_{\bf x}\la\vec{ \sigma}_{\bf 0}\cdot \vec{ \sigma}_{\bf x}\ra
\end{equation}
and enters into the two-point correlation function as
\begin{equation}
\beta\la\vec{ \sigma}_{\bf 0}\cdot \vec{ \sigma}_{\bf x}\ra=N\int_{-\pi}^{\pi}\frac{d^2 p}{(2\pi)^2}\frac{\exp(-i{\bf p}\cdot{\bf n})}{\chi^{-1}+Z^{-1}p^2+O(p^4)}.
\label{chi1}
\end{equation}
The susceptibility $\chi$ describes the magnetic response of the system when infinitesimal external fields are applied.  Actually, from comparison of (\ref{mass}) and (\ref{chi1}), it is clear that the limit $\chi\to \infty$ means the divergence of the dominant  length scale and corresponds to the continuum limit, thus $\chi$ playing a similar role with $1/M$.     In addition, the computation of the non-perturbative constant $C_{\chi}$ associated with $\chi$ itself is also of physical interest (the definition of $C_{\chi}$ will be given in the next section).

The purpose of the present paper is to apply the same approach taken in \cite{yam1} to the susceptibility of $O(N)$ sigma model on square lattice and examine the predictive power of Pad\'e-Borel transform on the strong coupling series for the approximation of continuum limit and non-perturbative quantity $C_{\chi}$.   In the present paper, we confine ourselves with $N\ge 3$.

In the next section, we survey the series expansion at weak and strong couplings.  Then we make attempt to exploit strong coupling series to access the scaling behavior and compute the value of $C_{\chi}$.   For the purpose we use Pad\'e-Borel approximation method.   Some remarks on the roles played by $\beta=1/g^2$ and $\chi$ are also given.  Conclusion is given in the last section.

\section{Series expansions at weak and strong couplings}
The standard action of the two-dimensional lattice non-linear sigma model with $O(N)$ symmetry reads
\begin{equation}
S=-\beta\sum_{\bf n}\sum_{\mu=1,2}\vec{\s}_{\bf n}\cdot\vec{\s}_{\bf n+\bf e_{\mu}},
\label{sigmaaction}
\end{equation}
where ${\bf e}_{1}=(1,0),\,{\bf e}_{2}=(0,1)$ and
\begin{equation}
\beta=\frac{1}{g^2}.
\end{equation}
The vector $\vec{\sigma}=(\sigma_{1},\sigma_{2},\cdots, \sigma_{N})$ is constrained to satisfy at every sites, $\vec{\s}^2=N$.

From perturbative renormalization group at $N\ge 3$, $\chi$ near the continuum limit is found to behave as
\begin{eqnarray}
\chi&\sim& \beta C_{\chi}\exp(\frac{4\pi N\beta}{N-2})\Big(\frac{2\pi N\beta}{N-2}\Big)^{-(N+1)/(N-2)}\nonumber\\
& &\times \Big(1+\frac{b_{1}}{\beta}+\frac{b_{2}}{\beta^2}+\cdots\Big),
\label{chi}
\end{eqnarray}
where $b_{1}$ and $b_{2}$ represent the three- and four-loop contributions, respectively, and obtained in  \cite{fal,col,col2,shin} as
\begin{eqnarray}
b_{1}&=&\frac{1}{N(N-2)}(-0.1888+0.0626N),\nonumber\\
b_{2}&=&\frac{1}{N^2(N-2)^2}\nonumber\\
& &\times(0.1316+0.0187N-0.0202N^2-0.0108N^3).
\end{eqnarray}
The multiplied constant $C_{\chi}$ is not analytically known.  Only its large $N$ expansion to the first order is analytically obtained as \cite{col2}
\begin{equation}
C_{\chi}=\frac{\pi}{16}\Big(1-\frac{4.267}{N}+O(N^{-2})\Big).
\label{1/n}
\end{equation}
The leading $1/N$ correction to $C_{\chi}$ is big and the value $1/(16\pi)$ in the large $N$ limit may not become an approximation for moderate values of $N$.

At strong coupling the susceptibility is expanded in powers of $\beta$.  Butera and Comi \cite{butera} obtained $\chi$ up to $\beta^{21}$.  To several orders $\chi$ is written as
\begin{eqnarray}
\chi&=&\beta\Big(1+4\beta+12\beta^2+\frac{72+32N}{N+2}\beta^3+\frac{200+6N}{N+2}\beta^4\nonumber\\
& &+\frac{8(284+147N+20N^2)}{(N+2)(N+4)}\beta^5+\cdots\Big)
\label{chibeta}
\end{eqnarray}
The result of inversion then reads
\begin{eqnarray}
\beta&=&\chi\Big(1-4\chi+20\chi^2-\frac{8(29+14N)}{2+N}\chi^3\nonumber\\
 & &+\frac{4(374+169N)}{2+N}\chi^4\nonumber\\
& &-\frac{8(5212+3451N+538N^2)}{(2+N)(4+N)}\chi^5+\cdots\Big).
\label{betachi}
\end{eqnarray}
To the $21$st order, the sign of coefficients of the series (\ref{betachi}) is alternative for all $N\ge 0$.   
  Based upon the above series effective at large lattice spacing, we attempt to recover the scaling behavior of $\beta$ and then estimate the non-perturbative constant $C_{\chi}$ for $N\ge 3$.

\section{Analysis by the use of Pad\'e-Borel method}
In this work, Borel transform with respect to $z$ has the meaning generalized to act not only on the series $\sum \frac{a_{n}}{z^n}$ but also on logarithms $z^{-const}(\log z)^k,\,\,(k={\rm integer})$, $(\log z)^{l}[\log(\log z)]^{m}\,\,(k,l={\rm integer})$, exponential functions and so on.  The transformation is carried out by taking a certain limit in delta expansion \cite{yam2}.  In some cases, it is also useful to exploit following integral representation,
\begin{equation}
{\cal B}[f]=\int_{C}\frac{dz}{2\pi i}\,\frac{e^{\zeta z}}{z}f(z) ,\quad \zeta=\frac{1}{\bar z}
\end{equation}
where $\zeta^{-1}=\bar z$ denotes the Borel counter part of $z$.  The contour $C$ of the integral is along with the straight line parallel to the imaginary axis on the complex $z$ plane.   All the non-analytic ingredients such as poles and cuts should be  in the left side of the contour if they are. 

To study the behavior of $\beta$ near the continuum limit we turn to the Borel transformed counter part, ${\cal B}[\beta]=\bar\beta$, at small and large $\bar\chi$ where $\bar\chi$ represents the Borel counter part of $\chi$.  The Borel transformed quantities have no direct physical meaning.  But they have inherited from physical quantities information we like to know, such that the constant $C_{\chi}$, logarithmic behavior with the coefficient $(4\pi)^{-1}$ and so on.

The Borel transform of $\beta$ at small $\chi$ reads
\begin{eqnarray}
\bar\beta&=&\bar\chi(1-\frac{4}{2!}\bar\chi+\frac{20}{3!}\bar\chi^2-\frac{8(29+14N)}{(2+N)4!}\bar\chi^3\nonumber\\
& &+\frac{4(374+169N)}{(2+N)5!}\bar\chi^4\nonumber\\
& &-\frac{8(5212+3451N+538N^2)}{(2+N)(4+N)6!}\bar\chi^5+\cdots).
\label{borelbeta}
\end{eqnarray}
To compare the Pad\'e approximants of above with the perturbative results, we use the following four-loop result,
\begin{eqnarray}
\beta&\sim& \frac{N-2}{4\pi N}\bigg\{t+\frac{3}{N-2}\log\Big[\frac{1}{2}t\Big]\nonumber\\
& &
+\frac{-2(N-2)^2 b_{1}+9\log[\frac{1}{2}t]}{(N-2)^2 t}\nonumber\\
& &+\frac{1}{(N-2)^3 t^2}\Big(-2(N-2)^2b_{1}+(2b_{1}^2-4b_{2})(N-2)^3\nonumber\\
& &+(27+6(N-2)^2b_{1})\log[\frac{1}{2}t]-\frac{27}{2}\log[\frac{1}{2}t]^2\Big)\bigg\},
\label{contbeta}
\end{eqnarray}
where
\begin{equation}
t=\log\Big[\frac{\chi}{C_{\chi}(\frac{N-2}{2\pi N})}\Big].
\end{equation}
The result of Borel transform of each term in (\ref{contbeta}) reads
\begin{eqnarray}
{\cal B}[\log \chi]&=&\log \bar \chi+\gamma_{E},\\
{\cal B}[\log\log \chi]&=&\log(\log \bar \chi+\gamma_{E})+\frac{\zeta(2)}{2(\log \bar \chi+\gamma_{E})^2}\nonumber\\
& &+O((\log \bar \chi)^{-3}),\\
{\cal B}[\frac{1}{\log \chi}]&=&\frac{1}{\log \bar \chi+\gamma_{E}}+O((\log \bar \chi)^{-3}),\\
{\cal B}[\frac{\log\log \chi}{\log \chi}]&=&\frac{\log(\log \bar \chi+\gamma_{E})}{\log \bar \chi+\gamma_{E}}+O((\log \bar \chi)^{-3}),\\
{\cal B}[\frac{1}{(\log \chi)^2}]&=&\frac{1}{(\log \bar \chi+\gamma_{E})^2}+O((\log \bar \chi)^{-3}),\\
{\cal B}[\frac{\log\log \chi}{(\log \chi)^2}]&=&\frac{\log(\log \bar \chi+\gamma_{E})}{(\log \bar \chi+\gamma_{E})^2}+O((\log \bar \chi)^{-3}),\\
{\cal B}[\frac{(\log\log \chi)^2}{(\log \chi)^2}]&=&\frac{(\log(\log \bar \chi+\gamma_{E}))^2}{(\log \bar \chi+\gamma_{E})^2}+O((\log \bar \chi)^{-3}),
\end{eqnarray}
where $\gamma_{E}=0.577216\cdots$.  Then, we arrive at 
\begin{eqnarray}
\bar \beta&\sim& \frac{N-2}{4\pi N}\bigg\{\bar t+\frac{3}{N-2}\Big(\log\Big[\frac{1}{2}\bar t\Big]+\frac{\zeta(2)}{2\bar t^2}\Big)\nonumber\\
& &+\frac{-2(N-2)^2 b_{1}+9\log[\frac{1}{2}\bar t]}{(N-2)^2\bar t}\nonumber\\
& &+\frac{1}{(N-2)^3\bar t^2}\Big(-2(N-2)^2b_{1}+(2b_{1}^2-4b_{2})(N-2)^3\nonumber\\
& &+(27+6(N-2)^2b_{1})\log[\frac{1}{2}\bar t]-\frac{27}{2}\log[\frac{1}{2}\bar t]^2\Big)\bigg\}\nonumber\\
&=&\bar\beta_{cont}
\label{borel}
\end{eqnarray}
where 
\begin{equation}
\bar t=\log\Big[\frac{\bar\chi}{C_{\chi}(\frac{N-2}{2\pi N})}\Big]+\gamma_{E}.
\end{equation}

Though  the transformed series (\ref{borelbeta}) in $\bar\chi$ is an entire series, it remains to show slow convergence to the exact result.  To remedy the situation, we further make use of Pad\'e method to extrapolate the transformed series to the large $\bar\chi$ region.   Since $\bar\beta$ behaves logarithmically for large $\bar\chi$, it would be natural to exploit the diagonal type of Pad\'e approximants, $\bar\beta_{[m/m]}$.  Here, $\bar\beta_{[m/m]}$ denotes a rational function of $\bar \chi$, where both of numerator and denominator are polynomials of degree $m$.  Then we will compare the behavior of $\bar\beta_{[m/m]}$ with $\bar\beta_{cont}$ to the four-loop level.  From $N=3$ to $20$ and at $N=\infty$, we have constructed diagonal Pad\'e approximants at $6$th, $8$th, $10$th $\cdots$, $22$nd orders.  The $22$nd order is the highest order available from the work of Butera and Comi \cite{butera}.  Singularity at positive real axis of $\bar\chi$ appears at $20$th order for $N=3$, $4$, $\cdots$, $16$.  In other cases, there is no pole at $\bar \chi>0$.   Thus the occurrence of the singularity at positive real axis is presumably accidental, and we assume that the singularity is originally absent on the positive real axis at all $N\le 3$.
  In any case, we show here the quantitative results at $22$nd order where the pole is absent on the positive real axis.

To see how Pad\'e approximants recover the continuum scaling, we have plotted $\bar\beta_{[11/11]}$ and $\bar\beta_{cont}$ for $N=3,4,8$ and $\infty$ ( $\bar\beta_{cont}$ is drawn by using estimated value $C_{app}$ for $C_{\chi}$.).  From Fig. 1, we observe the approximate continuum scaling at $N=8$ and $N=\infty$ already around $\log\bar\chi\sim 2$.   On the contrary, it may be hard to say that the scaling is seen at $N=3$ and perhaps at $N=4$.   Though $N=3$ and $4$ cases lack a clear signal of scaling, we have carried out estimation of $C_{\chi}$ at all $N\ge 3$ by the fitting of $\bar\beta_{cont}$ to $\bar\beta_{[m/m]}$.    In the fitting, recall that there is one unknown quantity $C_{\chi}$ which we like to estimate.    We vary the value of $C_{\chi}$ and seek the value at which the curve $\bar\beta_{cont}$ is tangent at a point  to the Pad\'e approximant.  We show in Table1 the estimated result denoted by $C_{app}$ at $22$nd order from $N=3$ to $N=20$.    Our results are slightly smaller than those of Butera and Comi \cite{butera}. 
  The point around which the fitting is realized is $\bar\chi=24.850$, $14.797$, $11.903$, $10.558$, $9.821$, $ 9.644$, $\cdots$, respectively for $N=3$, $4$, $5$, $6$, $7$, $8$, $\cdots$.
\begin{figure}[h]
\centering
\includegraphics[scale=0.75]{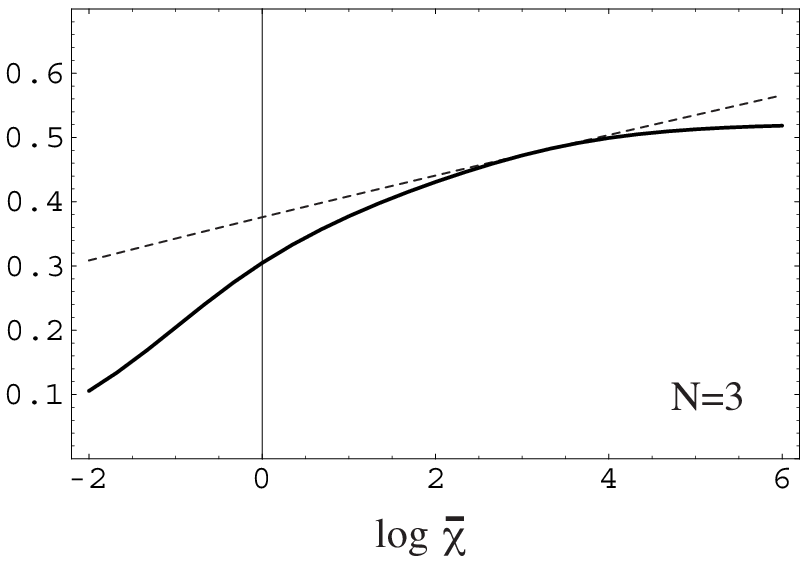}
\includegraphics[scale=0.75]{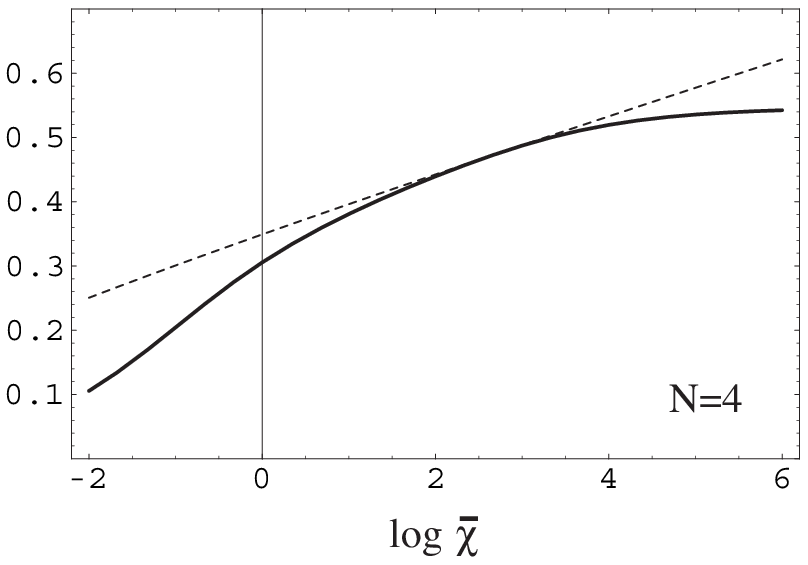}
\includegraphics[scale=0.75]{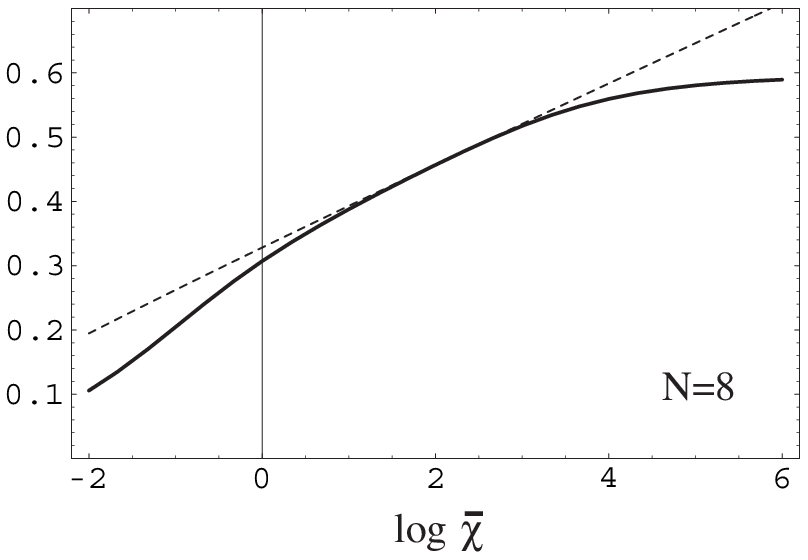}
\includegraphics[scale=0.75]{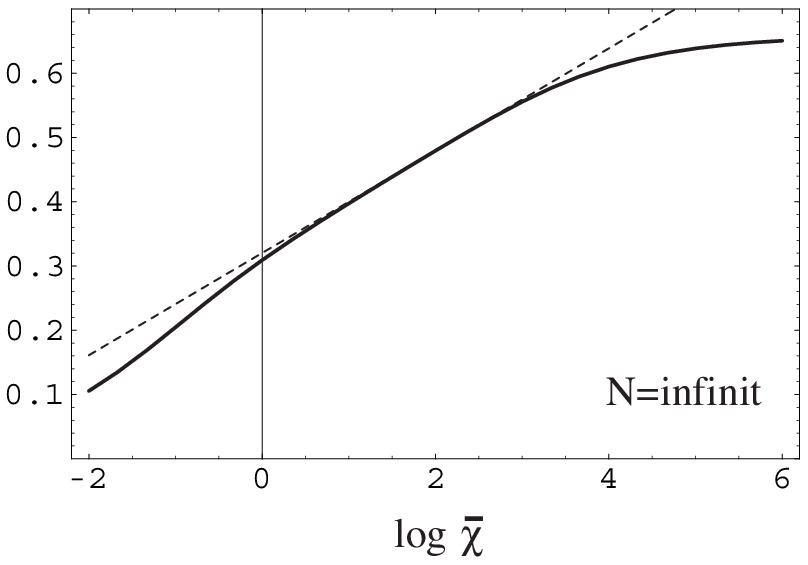}
\caption{ First three plots show four-loop perturbative $\bar\beta$ (dotted lines) and Pad\'e -Borel approximants at order $[11/11]$, respectively for $N=3,\,4$ and $8$ (solid lines).    The last plot shows $\bar\beta$ at $N=\infty$ in the continuum limit (dotted line) and Pad\'e -Borel approximants at order $[11/11]$ (solid line).}
\label{fig:1}  
\end{figure}

\begin{table}[h]
\caption{
Estimated values denoted as $C_{app}$ for the non-perturbative constant $C_{\chi}$.    On estimation we have used Pad\'e-Borel approximants of $\bar\beta$ at order $[11/11]$.     $C_{BT}$ means the estimation due to Butera and Comi \cite{butera}. } 
\label{tab:1}  
\begin{center}
\begin{tabular}{cccccc}
\hline\noalign{\smallskip}
$ N $ & $  C_{app}$ & $  C_{BT}$ & $ N$ & $ C_{app} $ & $  C_{BT}$ \\
\noalign{\smallskip}\hline\noalign{\smallskip}
$ 3$ &   $   0.00932 $ & $       $ & $\quad  12$ & $  0.13361   $& $   0.134 $  \\ 
$ 4$ & $  0.03273 $ & $   0.034 $   & $\quad  13$ & $   0.13782    $& $           $  \\ 
$5$ & $   0.05611  $  & $   0.059 $    & $\quad 14$ & $  0.14204    $& $   0.143 $ \\ 
$6$ &$  0.07526  $   & $   0.077 $    & $\quad 15$ & $   0.14574    $& $     $ \\ 
$ 7$ & $  0.09046  $  & $               $   & $\quad 16$ & $   0.14899   $  \\ 
$8$ & $   0.10258   $ & $   0.1035 $    & $\quad 17$ & $   0.15187   $  \\ 
$ 9$ &$   0.11304  $  & $            $    & $\quad 18$ & $  0.15446    $\\ 
$10$ &$  0.12048    $ & $   0.1212 $     & $\quad 19$ & $ 0.15678    $\\ 
$11$ &$  0.12723   $  & $     $    & $\quad 20$ & $   0.15886    $\\
\noalign{\smallskip}\hline
\end{tabular}   
\end{center}
\end{table}

Comparison with Monte Carlo results is available for $N=3$, $4$ and $8$.  At $N=3$, the two results, $C_{\chi}=0.0146(\pm 0.0010)$ \cite{cara} and $C_{\chi}=0.0130(\pm 0.0005)$ \cite{all} were reported.  
  At $N=4$, $C_{\chi}=0.0329(\pm 0.0016)$ in \cite{ed} and $C_{\chi}=0.0383(\pm 0.0010)$ in \cite{cara}.  At $N=8$, $C_{\chi}=0.1037(\pm 0.0004)$ in \cite{cara} and $C_{\chi}=0.1028(\pm 0.0002)$ in \cite{all}.      Apparently our estimation at $N=3$ is much smaller than any of Monte Carlo data.   At $N=8$ on the other hand, our result is consistent with the Monte Carlo results.

One of sources of discrepancy is the cut-off effects at weak coupling.  Formal expansion of the standard action yields terms irrelevant in the continuum limit and these terms obscure the asymptotic scaling.   For example, according to \cite{balog}, the leading correction to the asymptotic scaling of perturbative renormalization group result (\ref{contbeta}) becomes larger for smaller $N$.  This may explain the large discrepancy at $N=3$.  Unfortunately, since concrete information is not available yet, we cannot continue further.   In the large $N$ limit, however, we can discuss on the issue quantitatively.  The discussion will be presented later.  

Now, we turn to show that one can gain rough range of uncertainty of the estimation by the investigation of the near diagonal Pad\'e-Borel approximants.
  As previously stated, known logarithmic behavior of $\bar\beta$ at large enough $\bar\chi$ selects the diagonal Pad\'e as the most appropriate one.  However, reliable approximants would have stability for small shift of degrees of numerator and denominator.   Fig. 2 shows the plot of three graphs of $\bar\beta$ for $N=3$ at $22$nd order, $\bar\beta_{[12/10]}$, $\bar\beta_{[11/11]}$ and $\bar\beta_{[10/12]}$.  For $\bar\beta_{[12/10]}$ and $\bar\beta_{[10/12]}$, clear sign of limitation around $\log \bar\chi\sim 2$ is indicated.   The fitting gives $C_{\chi}=0.01355$ for $\bar\beta_{[12/10]}$ and $C_{\chi}=0.01448$ for $\bar\beta_{[10/12]}$.   We have examined the results of near diagonal Pad\'e at $22$nd order from $N=3$ to $10$.  The results are summarized in Table 2.

\begin{figure}[h]
\centering
\includegraphics[scale=0.75]{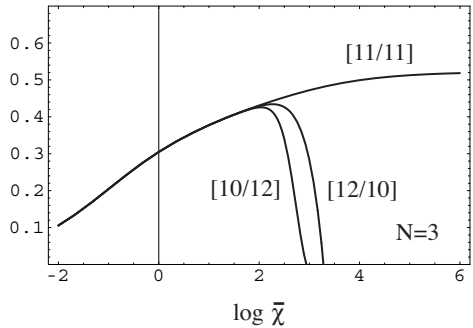}
\caption{Three graphs of $\bar\beta$ for $N=3$ at $22$nd order, $\bar\beta_{[12/10]}$, $\bar\beta_{[11/11]}$ and $\bar\beta_{[10/12]}$.}
\label{fig:2}  
\end{figure}

\begin{table}[h]
\caption{
Estimated results of $C_{\chi}$ at $22$nd order for approximants at diagonal and near diagonal assignment of numerator- and denominator-degrees.  The last column shows the ratio of the average of the two off-diagonals to the diagonal.} 
\label{tab:2}  
\begin{center}
\begin{tabular}{ccccc}
\hline\noalign{\smallskip}
$ N $ & $[12/10]$ & $[11/11]$ & $[10/12]$ & $ratio$  \\
\noalign{\smallskip}\hline\noalign{\smallskip}
$3$ &   $   0.01355 $ & $0.00932$ & $0.01448$ & $1.504$   \\ 
$4$ & $  0.03629$ & $0.03273$ & $0.03703$ & $1.120$  \\ 
$5$ & $  0.05899  $ & $0.05611$ & $0.05962$ & $1.057$  \\ 
$6$ & $  0.07770  $ & $0.07526$ & $0.07822$  & $1.036$ \\ 
$7$ & $  0.09261  $ & $0.09046$ & $0.09311$ & $1.027$  \\ 
$8$ & $  0.10456  $ & $0.10258$ & $0.10504$ & $1.022$  \\ 
$9$ & $  0.11427  $ & $0.11240$ & $0.11472$ & $1.019$  \\ 
$10$ & $  0.12228  $ & $0.12048$ & $0.12274$  & $1.017$ \\ 
\noalign{\smallskip}\hline
\end{tabular}   
\end{center}
\end{table}
Note the large difference for small $N$ between $[12/10]$ and $[11/11]$, and $[10/12]$ and $[11/11]$.  For example, 
differences between $[12/10]$ and $[11/11]$ are $0.00423$, $0.00356$ and $0.00198$, respectively for $N=3$, $4$ and $8$.  In particular, for $N=3$, the size of $C_{\chi}$ itself is small and, as explicitly shown in Table 2,  the ratio of the average of $[12/10]$ and $[10/12]$ to $[11/11]$ is about $1.5$.  Thus, we find that the estimation status for $N=3$ has not reached to enough stability yet.   We have also checked that the proliferation range of estimated values within $[m+1/m-1]$, $[m/m]$ and $[m-1/m+1]$ becomes narrower as the order $2m$ increases ( see Table 2).   These features would lead us to conclude that, for $N=3$, the order of strong coupling expansion is still short.

So far, we have discussed the model at finite $N$.  We now turn to the $N$ limit.  At $N=\infty$,  we find the exact result from (\ref{1/n}),
\begin{equation}
C_{\chi}(\infty)=\frac{\pi}{16}=0.19635\cdots.
\label{limit}
\end{equation}
Our result at $22$nd order with diagonal $\bar\beta_{[11/11]}$ gives $C_{\chi}=0.19970\cdots$ which is $1.7$ percents larger than (\ref{limit}).  
One reason of the discrepancy is, of course, the truncation of large $\bar\chi$ series.  However, the presence of the lattice artifact also prevents us from accurate estimation.  As is well known, $\chi$ dependence of $\beta$ is exactly specified by the gap equation,
\begin{equation}
\beta=\int_{-\pi}^{\pi}\frac{d^2 p}{(2\pi)^2}\frac{1}{\chi^{-1}+2\sum_{\mu=1,2}(1-\cos p_{\mu})}.
\label{gap}
\end{equation}
On the lattice artifact, we can estimate its effect as follows.  Near the continuum limit, we have series expansion,
\begin{eqnarray}
\beta&=&\frac{1}{4\pi}\log(32\chi)+\frac{1}{32\pi\chi}(-\log(32\chi)+1)\nonumber\\
& &+\frac{1}{1024\pi\chi^2}(5\log(32\chi)-7)+\cdots.
\label{artifact}
\end{eqnarray}
The second, third and higher order terms represent the lattice artifacts.   These are neglected in (\ref{chi}) and (\ref{contbeta}).  Borel transform of (\ref{artifact}) reduces the effect to some extent as
\begin{equation}
\bar\beta=\frac{1}{4\pi}(\log(32\bar\chi)+\gamma_{E})-\frac{1}{32\pi\bar \chi}-\frac{5}{1024\pi\bar\chi^2}+\cdots.
\label{artifact2}
\end{equation}
If we include the second term $-\frac{1}{32\pi\bar\chi}$ we have $C_{\chi}=0.19652$ by $\bar\beta_{[11/11]}$ at $\bar\chi=6.331$.  Third term incorporation gives $C_{\chi}=0.19639$ at $\bar\chi=5.69$.   Thus, much accurate estimation is obtained by fitting around a non-large value of $\bar\chi$. 

As the last argument, let us consider what comes out when $\chi$ is expressed in $\beta$ as in the conventional manner.   From the perturbative result (\ref{chi}), it follows that 
\begin{eqnarray}
\log\frac{\chi}{C_{\chi}}&\sim& \frac{4\pi N\beta}{N-2}-\frac{N+1}{N-2}\log\frac{2\pi N}{N-2}-\frac{3}{N-2}\log\beta\nonumber\\
& &+\frac{b_{1}}{\beta}+\frac{2b_{2}-b_{1}^2}{2\beta^2}+\cdots.
\label{chi2}
\end{eqnarray}
When taking $\beta$ as the basic variable by which $\chi$ is controlled, we must consider the $\beta\to \infty$ limit.  Then it would be nice if the strong coupling series in $\beta$ could be Borel transformed and the result allows us approximation of scaling.
The right-hand side of (\ref{chi2}) seems to be convenient form for Borel transforming with respect to $\beta^{-1}=g^2$.  However, the scheme does not work well compared with the inverted version which we have presented.  We like to clarify why it is better to express $\beta$ as a function of $\chi$ than its reverse, when Borel transform is exploited.  We focus on the large $N$ limit since in this case all of necessary things are computed exactly (Note that, in the $N\to \infty$ limit, $\chi=\frac{1}{M}$ and the following argument exactly applies also to $M$ vs $\beta$).

First let us remind that, by taking the $N\to \infty$ limit of (\ref{chi2}), we find that only the first two terms survives in (\ref{chi2}), resulting $\log(\chi/C_{\chi})\to 4\pi\beta-\log(2\pi)$.  On the contrary, we find from  (\ref{artifact}) that $
32\chi=e^{4\pi\beta}+4(4\pi\beta-1)+4(-13+12\pi\beta-32\pi^2\beta^2)e^{-4\pi\beta}+O(e^{-8\pi\beta})$ and
\begin{eqnarray}
\log\frac{\chi}{C_{\chi}}&\sim& 4\pi\beta-\log(2\pi)+4(4\pi\beta-1)e^{-4\pi\beta}\label{chi3}\\
& &+4(-15+28\pi\beta-64\pi^2\beta^2)e^{-8\pi\beta}+O(e^{-12\pi\beta}).\nonumber
\end{eqnarray}
The terms of exponential in (\ref{chi3}) are the cut-off effects which rapidly vanish in the continuum limit.   The cut-off effects are not included in (\ref{chi2}), but they are enhanced by Borel transformation in the following way:  
Suppose that we make the Borel transform to access the large $\beta$ behavior of $\chi(\beta)$.   Typical example of Borel transform may be given by ${\cal B}[e^{-4\pi \beta}]$ (transform should be done with respect to $\beta^{-1}$),
\begin{equation}
{\cal B}[e^{-4\pi\beta}]=\sum_{n=0}^{\infty}\frac{(-4\pi\bar\beta)^n}{(n!)^2}=J_{0}\Big(2\sqrt{4\pi\bar\beta}\Big)
\end{equation}
where $J_{0}$ denotes the Bessel function.  It is also easy to see that ${\cal B}[\beta e^{-4\pi\beta}]=\int_{0}^{\bar\beta}J_{0}(2\sqrt{4\pi z})dz$.   The term $\beta^{k} e^{-8\pi\beta}$ $(k=0,1,2)$ transforms in the similar manner and the result ends with oscillatory function.  
${\cal B}[\log\chi]$ at large $\bar\beta$ is also oscillatory and the amplitude is not small.  Thus, ${\cal B}[\log\chi]$ in strong coupling expansion faithfully recovers that oscillatory behavior and is inadequate for the examination of continuum scaling and estimation of the non-perturbative constant $C_{\chi}$.  Rather, since the lattice artifact is exponentially small, it is better to use directly the Pad\'e method on original strong coupling series, for example, as performed in \cite{butera}.

\section{Conclusion}
We have investigated the application of Pad\'e-Borel method to non-linear $O(N)$ sigma model at $N\ge 3$.   By expanding the bare coupling in terms of susceptibility $\chi$ at large lattice spacings, we have examined the scaling behavior near the continuum limit by using Pad\'e-Borel method.   Scaling behavior has been observed at large enough $N$.   For $N=3$, we find large discrepancy between the estimated $C_{\chi}$ and its Monte Carlo data.   For $N=4$ the estimated value of $C_{\chi}$ is close to existing Monte Carlo data and the result in \cite{butera}.    For larger $N$, scaling behavior becomes gradually clearer as $N$ increases and the estimated values of $C_{\chi}$ are in good agreement with those in literatures.  

As for small $N$ cases, it is desirable that the longer series is brought to scaling examination.  Though not on the square lattice, there exists a literature where longer series expansion is computed on honeycomb lattice \cite{camp}.   We hope to report the result elsewhere in the near future.

\end{document}